\documentstyle[twocolumn,aps,epsfig]{revtex} 
 
\draft 
\begin{document} 
 
 
\title{Muon-Spin Rotation Measurements of the Magnetic Field Dependence 
of the Vortex-Core Radius and Magnetic Penetration Depth in NbSe$_2$} 
 
\author{ J.~E.~Sonier, R.~F.~Kiefl, J.~H.~Brewer, 
J.~Chakhalian, S.~R.~Dunsiger, 
W.~A.~MacFarlane, R.~I.~Miller and A.~Wong}  

\address{TRIUMF, Canadian Institute for Advanced Research 
 and Department of Physics and Astronomy, University of British Columbia, 
 Vancouver, British Columbia, Canada V6T 1Z1 } 

\author{G.~M.~Luke} 
 
\address{Department of Physics, Columbia University, New York, 
New York 10027, U.S.A.}

\author{J.~W.~Brill}

\address{Department of Physics and Astronomy, University of Kentucky, 
Lexington, Kentucky 40506-0055, U.S.A.}

\date{March, 1997} 
\date{ \rule{2.5in}{0pt} } 
 
\maketitle 
\begin{abstract} \noindent 
Muon-spin rotation spectroscopy ($\mu$SR) has been used to
measure the internal magnetic
field distribution in NbSe$_2$ for 
$H_{c1} \! \ll \! H \! < \! 0.25~H_{c2}$. 
The deduced profiles of the supercurrent density
$J_s$ indicate that the vortex-core
radius $\rho_0$ in the bulk decreases sharply with increasing 
magnetic field. This effect, which is attributed to
increased vortex-vortex interactions, does not agree
with the dirty-limit microscopic theory.
A simple phenomenological equation
in which $\rho_0$ depends on the intervortex spacing
is used to model this behaviour. 
In addition, we find for the first time that the in-plane magnetic 
penetration depth $\lambda_{ab}$ increases linearly with $H$ in 
the vortex state of a conventional superconductor.
\end{abstract} 
\begin{center}
\rule{0.3\textwidth}{0.25pt} 
\end{center}

Although the bulk of superconductivity research in the past decade has 
been dominated by the study of the high-$T_c$ cuprates, there remain
unresolved issues concerning conventional superconductors. 
In the Meissner state of a conventional $s$-wave superconductor,
application of a magnetic field $H$ induces pair-breaking and
the ensuing backflow of
quasiparticles gives rise to a nonlinear dependence
of induced supercurrent with respect to the superfluid
velocity \cite{Yip:92}. The solution of the corresponding
nonlinear London equation gives a penetration depth $\lambda$
which changes quadratically with $H$. This has been
confirmed experimentally in Sn, In \cite{Sridhar:86} and 
V$_3$Si \cite{Hanaguri:95}.
Theoretical predictions in the vortex state are less developed
and the $H$-dependence of $\lambda$ in the vortex state
of a conventional superconductor has never been determined experimentally.
Similarly, the $H$-dependence of the vortex-core radius $\rho_0$ 
has never been measured in the bulk
of a conventional superconductor. 
In a conventional $s$-wave superconductor 
$\rho_0 \! \sim \! \xi$ (Ref.~\cite{Tinkham:96}),
where $\xi$ is the coherence length.
Measurements of $\rho_0$ bear directly
on the electronic structure of an isolated vortex  
and provide a necessary comparison for ongoing studies on high-$T_c$
materials. 

NbSe$_2$ is particularly well suited for a $\mu$SR study 
of the vortex state since the
geometry of the vortex lattice is well 
established---thus removing one of the largest experimental uncertainties.
Scanning tunneling microscopy (STM) measurements at the surface
\cite{Hartmann:93,Hess:89,Hess:92} and small angle neutron scattering
(SANS) measurements in the bulk \cite{Gammel:94} have produced high
quality images of a nearly perfect triangular lattice with long
range order. Also the coherence length and the magnetic penetration
depth in NbSe$_2$ are nearly ideal for a $\mu$SR investigation
of the vortex cores. In particular, the large coherence length
and correspondingly small value of $H_{c2}$ ($< \! 4$~T) imply
a large signal from the vortex-core regions at moderate
magnetic fields. 

Most theoretical
calculations have modelled the core structure using 
GL theory, which is strictly valid only near the superconducting
phase boundary.
Recently the field dependence of the vortex-core
radius has been determined deep in the superconducting
state from the microscopic theory in the dirty limit, 
by Golubov and Hartmann \cite{Golubov:94}. 
The vortex-core radius was found to  
decrease monotonically with increasing applied magnetic field due
to the increased strength of the vortex-vortex interactions.
Although the authors reported good agreement with 
STM measurements \cite{Hartmann:93} at the surface of
NbSe$_2$ at $T \! = \! 0.6~T_c$, $\rho_0$ was somewhat arbitrarily defined 
and the uncertainty in the measurements was large.  
The results were relatively surprising since 
NbSe$_2$ is a clean 
superconductor---the ratio of the coherence length to the mean free
path in the $\hat{a}$-$\hat{b}$ plane is
$\xi_0 / l \! \sim \! 0.15$ \cite{Takita:85}.
In the present study, we make a comparison
between the vortex-core radius measured 
by $\mu$SR in the bulk of NbSe$_2$ and that determined from 
the spatial dependence of the supercurrent density $J_s$ in
the dirty-limit microscopic theory. 
The results show that the dirty-limit approximation
is invalid, even at $T \! = \! 0.6~T_c$. The $H$-dependence of
$\rho_0$ is fit to a simple phenomenological
equation arising from our observation that both
$\lambda_{ab}$ and the GL parameter $\kappa$ vary linearly
with magnetic field.
We also report for the first time, $\mu$SR measurements of the
$H$-dependence of $\lambda_{ab}$ in the vortex state of a
conventional superconductor. Contrary to the $H^2$ dependence
in the Meissner state, $\lambda_{ab}$
shows a strong linear $H$-dependence in NbSe$_2$
over a significant range of applied field.

In a $\mu$SR experiment the implanted muon samples the 
distribution of local magnetic
fields in the bulk of a type-II superconductor in the vortex state. 
One monitors the ensemble averaged muon-spin precession
signal. The frequency of 
precession for any given muon is directly proportional 
to the local magnetic field at the muon site. 
Further details regarding the $\mu$SR technique may be found
elsewhere (see for example Ref.~\cite{Riseman:95}). 
The present $\mu$SR study of NbSe$_2$ was performed on
the M20 beam line at TRIUMF. 
The single crystal of NbSe$_2$ used in this experiment had a mass of
43~mg and a surface area of $\sim$30~mm$^2$. The superconducting transition
$T_c$ and $H_{c2} (T \! = \! 0)$ determined from magnetization were $7.0$~K 
and $3.5$~T, respectively. For details regarding the sample growth
see Ref.~\cite{Drulis:91}.
The sample was mounted with its $\hat{c}$-axis
parallel to the applied field and beam direction.
$\mu$SR spectra were recorded under conditions of field cooling in a
$^4$He gas flow cryostat. A cup-shaped veto counter was used to
suppress unwanted background signal from muons which missed
the sample \cite{Schneider:93}.

In Fig.~1 the Fourier transforms of the
muon precession signal in NbSe$_2$ are shown for
different fields at $T \! = \! 0.33~T_c$. The real amplitude
of the Fourier transforms
is a good representation of the internal magnetic field distribution
from the vortex lattice convoluted with
small nuclear dipolar fields.      
Note the small peak near zero in the top panel of Fig.~1.
This is due to a small 
\begin{figure}[b]
\epsfig{figure=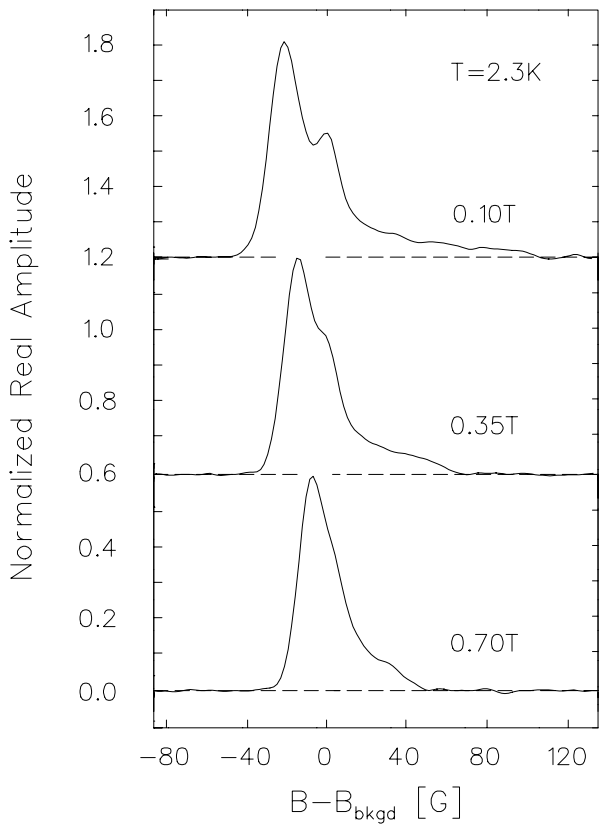, width=\linewidth}
\\
\caption[]{The Fourier transforms of the muon spin precession
signal in NbSe$_2$ after field cooling to $T \! = \! 0.33~T_c$ in 
magnetic fields of $H \! = \! $ 0.1, 0.35 and 0.7~T. The
average magnetic field of the residual background signal is denoted
as $B_{bkgd}$.}
\end{figure}
\noindent (2\%) residual background signal 
due to muons which miss the sample. 
Lineshapes have been renormalized to the same maximum amplitude.
The sharp features expected for a perfect triangular vortex lattice,
such as the Van Hove singularity at the saddle point,
are obscured partly by broadening effects of the finite Fourier transform
and by flux-line lattice disorder. 
Nevertheless there is a clear high-field cutoff observed in the $\mu$SR
lineshape originating from the finite size of the vortex cores.
The effect of increasing $H$ on the high-field
cutoff is clearly seen in Fig.~1. 
At all of the magnetic fields
the signal-to-noise ratio of the high-field tail is 
so large that one can
unambiguously extract the vortex-core radius.

In order to test the strength of the pinning forces on
the vortex lattice the sample was cooled
in an applied field of 0.5~T
to 2.3~K after which the field was decreased by 7.5~mT. In our previous
studies of YBa$_2$Cu$_3$O$_{7-\delta}$ \cite{Sonier:94} we
found under similar conditions that the background signal shifted
to a lower frequency in response to the change in the applied field, while
the signal corresponding to muons stopping in
the sample remained unchanged. In NbSe$_2$ both signals 
shifted---indicating that the vortex lattice is 
considerably more weakly pinned than in
YBa$_2$Cu$_3$O$_{7-\delta}$.

The $\mu$SR spectra were fit in the time domain 
where there are no complications associated
with fitting finite Fourier transforms \cite{Sonier:97}. The
distribution of muon precession frequencies from the
vortex lattice was modelled with a 
theoretical field distribution generated from a 
GL model \cite{Hao:91}. The local field 
at any point in the $\hat{a}$-%
$\hat{b}$ plane is given in a suitable approximation by \cite{Yaouanc:97},
\begin{eqnarray} 
\label{eq:Br} 
 B(\mbox{\boldmath $\rho$}) & = & B_0 (1-b^4)\sum_{ {\bf G} } 
 { e^{-i {\bf G} \cdot \mbox{\boldmath $\rho$} } 
 \,\, u \, K_1(u) 
 \over 
 \lambda_{ab}^2 G^2}, \eqnum{1a} \\
\nonumber \\
\mbox{with,} \;\;\;\;\;
u^2 & = & 2 \, \xi_{ab}^2 G^2 (1+b^4)[1-2b(1-b)^2]. \eqnum{1b}
\end{eqnarray} 
Here $B_0$ is the average magnetic field, 
{\bf G} are the reciprocal lattice vectors,
$b \! = \! B_{0}/B_{c_2}$,
$\xi_{ab}$ is the GL coherence length and $K_1(u)$ is a modified
Bessel function. The cutoff factor $u\,K_1(u)$ accounts for the
finite size of the vortex core---whereas in the London model 
$B(\rho)$ diverges logarithmically as $\rho \! \rightarrow \! 0$. Recently,
Yaouanc {\it et al.} \cite{Yaouanc:97} showed that $u\,K_1(u)$
is a good approximation of the cutoff factor determined from
the exact numerical solutions of the GL equations \cite{Brandt:97}
at low reduced fields $b$.

The theoretical muon polarization function was generated by assuming
the field profile of Eq.~(\ref{eq:Br}) and then multiplying
by a Gaussian relaxation function $e^{-\sigma^2 t^2/2}$ to take
into account any residual disorder in the flux-line lattice 
and the contribution
of the nuclear dipolar moments
to the internal field distribution. 
The residual background signal was fit assuming a Gaussian broadened
distribution of fields.
All fitted parameters were treated
as independent variables.
From the fitted values of $\sigma$, 
the RMS deviation of the vortices from their ideal positions in the
triangular lattice was determined to be less than 3\% of the intervortex
spacing over the entire 
\begin{figure}[t]
\epsfig{figure=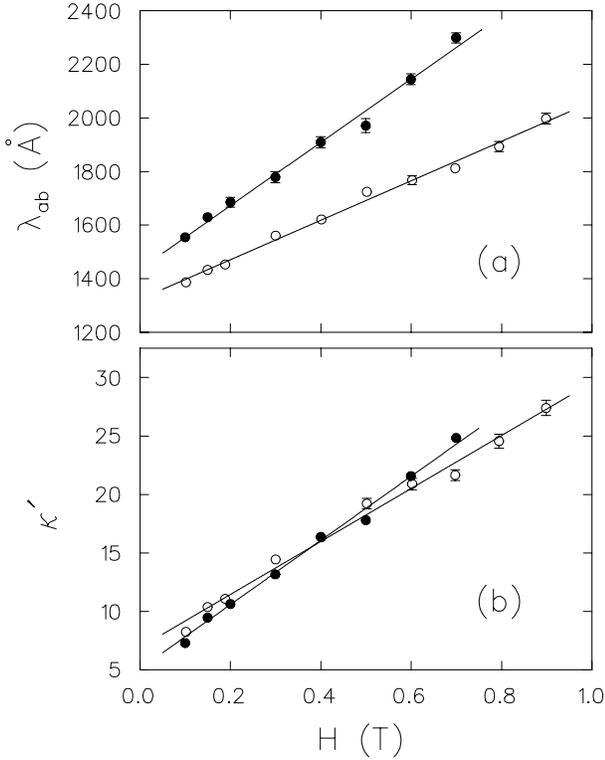, width=\linewidth}
\\
\caption[]{The magnetic field dependence of (a) $\lambda_{ab} (H)$
and (b) $\kappa^\prime(H) = \lambda_{ab} (H)/\rho_0(H)$ in the 
vortex state of NbSe$_2$ at
$T \! = \! 0.33~T_c$ (open circles) and $T \! = \! 0.6~T_c$ (solid circles).  
The solid line fits are described in the text.}
\end{figure}
\noindent field range studied. This small disorder is
consistent with STM and SANS imaging experiments on NbSe$_2$.

The magnetic field dependence of $\lambda_{ab}$ is
shown in Fig.~2(a). 
Contrary to the Meissner state, a linear-$H$
dependence is observed in the field range studied.
A fit to the linear relation
$\lambda_{ab}(H) \! = \! \lambda_{ab}(0) [ 1 + \! \beta h]$,
where $h \! = \! H/H_{c2}(T)$, gives
$\lambda_{ab} (0) \! = \! 1323$~\AA\ and $\beta \! = \! 1.61$ 
at $T \! = \! 0.33~T_c$ and $\lambda_{ab} (0) \! = \! 1436$~\AA\ and 
$\beta \! = \! 1.56$ at $T \! = \! 0.6~T_c$.
We note that $d[\Delta \lambda/\lambda(0)]/d(H/H_{c2})$
is considerably weaker than for 
YBa$_2$Cu$_3$O$_{6.95}$ (Ref.~\cite{Sonier:97}),
in which there is strong evidence for line nodes in
the superconducting energy gap function.

We define an effective vortex-core radius $\rho_0$ to be the distance
from the vortex center for which the supercurrent density
$J_s(\rho)$ reaches its maximum value. $J_s(\rho)$ was
obtained from fits of the data to Eq.~(\ref{eq:Br}) and the Maxwell relation
${\bf J}(\mbox{\boldmath $\rho$}) \! = \!
\mbox{\boldmath $\nabla$} \! \times \! {\bf B}(\mbox{\boldmath $\rho$})$.
In Fig.~3 $\mu$SR measurements of $\rho_0$ 
are shown as a function of $H$ at $T \! = \! 0.33$ (open circles) and $0.6~T_c$
(solid circles), along with the 
STM measurements (open squares) of Ref.~\cite{Hartmann:93}.
The smaller error bars and reduced scatter in the $\mu$SR data
reflects the statistical improvement of a $\mu$SR experiment which samples
a large number of vortices in the bulk of the crystal, 
as opposed to STM which averages the radius of a few
vortices at the surface.     
The dashed curve drawn through the STM results comes
from tunneling current $I(\rho)$ profiles calculated from the
Usadel equations, as explained in Ref.~\cite{Golubov:94}.
To generate $J_s(\rho)$ profiles from Usadel's dirty-limit theory
we extend the work of Ref.~\cite{Golubov:94} to include
the self-consistency equation for the vector potential ${\bf A}(\rho)$.
In cylindrical coordinates the equation of motion is \cite{Kramer:74},
\begin{equation} 
\frac{1}{\rho}\frac{d}{d\rho}\left(\rho \frac{d\theta}{d\rho}\right)
= \bar{\kappa}^{-2}A^2 \sin \theta \cos \theta - \Delta \cos \theta
+ \omega \sin \theta, \eqnum{2}
\label{eq:motion}
\end{equation}
where $\theta$ parametrizes Usadel's normal ($G \! = \! \cos \theta$)
and anomalous ($F \! = \! \sin \theta$) Green's functions \cite{Usadel:70},
$\Delta(\rho)$ is the order parameter, 
$\omega \! = \! (T/T_c)(2l+1)$ is the Matsubara frequency and
$\bar{\kappa} \! = \! [4 \pi^5 / 7 \zeta(3)]^{1/2}\kappa$ (where
$\kappa \! = \! \lambda/\xi$).
Equation~(\ref{eq:motion}) is supplemented by the following
self-consistency equations,
\begin{eqnarray}
& \Delta \ln & (T/T_c) = -2(T/T_c) \sum_{\omega} \left[ \Delta/\omega
- \sin \theta \right] \eqnum{3}
\label{eqn:sc1}\\
& J_s (\rho & ) = \frac{d}{d\rho}\frac{1}{\rho}\left(\frac{d}{d\rho} 
\rho A \right) = 16 \pi \bar{\kappa}^{-2}(T/T_c) A \sum_{\omega} 
\sin^2 \theta, \eqnum{4}
\label{eqn:sc2}
\end{eqnarray}
and the boundary conditions for singly quantized vortices with a
Wigner-Seitz cell radius $\rho_s \! = \! (\Phi_0/\pi H)^{1/2}$,
\begin{eqnarray}
& \Delta (0) =  \theta(\omega,0) = 0,& \,\,\,\,\,\,\,\,
\Delta^\prime(\rho_s)  =  \theta^\prime(\omega, \rho_s)=0 \eqnum{5} \\
& A (\rho  \! \rightarrow \! 0) \rightarrow -\bar{\kappa}/\rho,& \,\,\,
\,\,\,\,\, A(\rho_s) = 0 \eqnum{6}.
\end{eqnarray}
Equation~(\ref{eq:motion}) subject to these boundary conditions
was solved numerically for $\theta(\omega, \rho)$ starting with
the initial trial potentials $\Delta (\rho) \! = \! \Delta_0 \tanh(\rho)$
and $A(\rho) \! = \! \bar{\kappa}\left(1/\rho - \rho/\rho_s^2 \right)$.
Improved values of $\Delta (\rho)$ and $A(\rho)$ were obtained 
by including the self-consistent conditions (\ref{eqn:sc1}) and
(\ref{eqn:sc2}). The parameter $\kappa$
in equations (\ref{eq:motion}) and (\ref{eqn:sc2}) was determined
from fits to the data. 
We found $\kappa$ to be nearly 
\begin{figure}[b]
\epsfig{figure=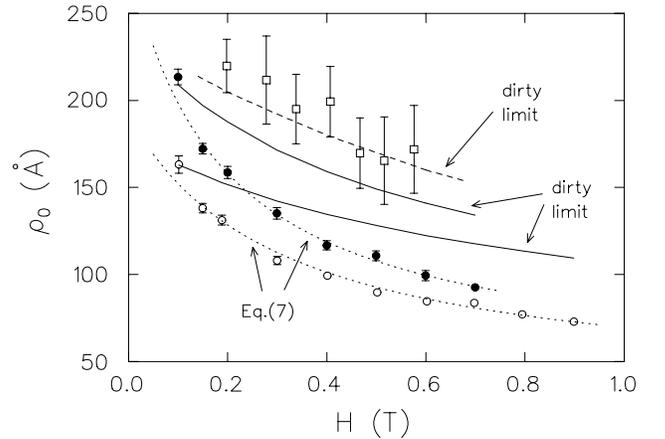, width=\linewidth}
\\
\caption[]{The magnetic field dependence of the vortex-core radius 
in NbSe$_2$ determined by STM \cite{Hartmann:93}
at $T \! = \! 0.6~T_c$ (open squares) and by $\mu$SR 
at $T \! = \! 0.33~T_c$ (open circles)
and $0.6~T_c$ (solid circles). The
solid curves are from calculations of supercurrent density
$J_s(\rho)$ profiles using $H_{c2}(0) \! = \! 3.5$~T, 
$H_{c2}(0.33~T_c) \! = \! 2.9$~T and $H_{c2}(0.6~T_c) \! = \! 1.9$~T 
from magnetization. The dotted curves through the
$\mu$SR data are from Eq.~(\ref{eqn:phenom}).}
\end{figure}
\noindent temperature independent
at all fields studied, which is consistent with the original
definition of $\kappa$ near $T_c$ in GL theory \cite{Ginzburg:50}.
The deduced values of $\rho_0$ were not very sensitive to $\kappa$. 
In Fig.~2(b) we define
$\kappa^\prime \! = \!  \lambda_{ab}/\rho_0 \!$,
where $\kappa^\prime (H) \! = \! 1.06 \kappa(H) - 1.98$ and 
$\kappa^\prime (H) \! = \! 1.12 \kappa (H) - 1.26$  
at $T \! = \! 0.33~T_c$ and $T \! = \! 0.6~T_c$, respectively. 
At both temperatures $\kappa^\prime$ (and hence $\kappa$) 
increases linearly with $H$. Fitting to the linear relation
$\kappa^\prime (H) \! = \! \kappa^\prime (0) [1 + \! \gamma h]$, we obtain
$\kappa^ \prime (0) \! = \! 6.9$ and $\gamma \! = \! 9.5$ 
at $T \! = \! 0.33~T_c$ 
and $\kappa^\prime (0) \! = \! 5.1$ and $\gamma \! = \! 10.2$ 
at $T \! = \! 0.6~T_c$. 

Figure~4. shows the theoretical $J_s (\rho)$ and
$\Delta (\rho)$ profiles together with the
$J_s (\rho)$ profile obtained from experiment
for a particular $T$ and $H$.  
The vortex-core radius taken from the $J_s (\rho)$ 
profiles of the dirty-limit theory are
shown as solid curves in Fig.~3. Not surprisingly
there is poor agreement with the $\mu$SR data 
at $T \! = \! 0.33~T_c$, where
thermal smearing of the bound states in the vortex core 
is negligible. 
Contrary to the STM results \cite{Hartmann:93}, however,
there is also poor agreement at $T \! = \! 0.6~T_c$---suggesting that the
dirty-limit theory does not adequately describe the shrinking of
the vortex-core radius with increasing $H$.
In the STM experiment it was necessary to arbitrarily define 
$\rho_0$ as the radius in which the tunneling current
decreased to $36\%$ of $I_{\rm max} \! - \! I_{\rm min}$ and 
$\Delta(\rho)/\Delta(\rho_s) \! = \! 1/\sqrt{2}$. However, we find that
the $J_s (\rho)$ profiles generated from the dirty-limit theory do not
peak exactly at a radius corresponding to $\Delta(\rho)/\Delta(\rho_s) \! = 
\! 1/\sqrt{2}$
for all values of $T/T_c$ and $H/H_{c2}$.
Thus our definition of $\rho_0$ should provide a better description 
of the true $H$-dependence of the vortex-core radius. 
The STM data may also be influenced somewhat by the
discontinuity in the energy spectrum of the vortex cores
which occurs at the sample surface. It has been suggested 
that the effect may be an enlargement of $\rho_0$ \cite{Klein:90}. 
\begin{figure}[b]
\epsfig{figure=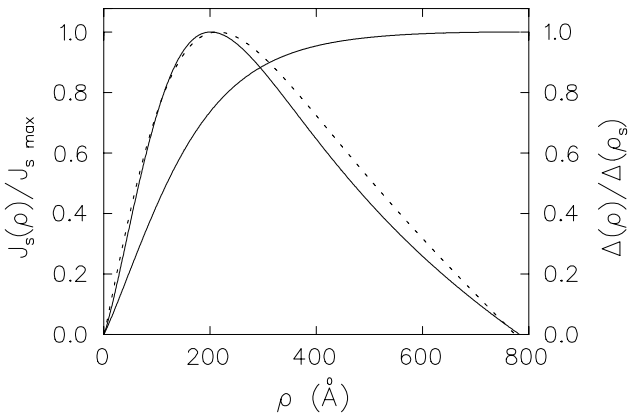, width=\linewidth}
\\
\caption[]{The current density and order parameter for $T \! = \! 0.6~T_c$ and
$H/H_{c2}(0.6~T_c) \! = \! 0.053$. The solid curves are from the dirty-limit
microscopic theory while the dashed curve is the supercurrent density from
a $\mu$SR measurement of the field distribution in NbSe$_2$.}
\end{figure}

On the other hand, the $\mu$SR results fit well
(see dotted curves in Fig.~3)
to the phenomenological equation,
\begin{eqnarray}
\rho_0(H) = \frac{\lambda_{ab}(H)}{\kappa^\prime (H)}
& = & \rho_0(0) \frac{[1+\beta h]}{[1+\gamma h]}, \eqnum{7}
\label{eqn:phenom} 
\end{eqnarray}
where $\rho_0 (0) \! = \! \lambda_{ab}(0)/\kappa^\prime (0)$.
From the fits to $\lambda_{ab}(H)$ and $\kappa^\prime (H)$,
$\rho_0(0) \! = \!  191$ and $282$~\AA\ at 
$T \! = \! 0.33$ and $0.6~T_c$, respectively.
For a triangular vortex lattice
$H \! \cong \! B_0 \! = \! 3 \Phi_0/ \sqrt{2} L^2$, where
$L$ is the intervortex spacing.
Thus for a given temperature 
Eq.~(\ref{eqn:phenom}) 
may be rewritten as a function of
the distance between vortices.

In conclusion, we have determined that the magnetic penetration depth
$\lambda_{ab}$ shows a linear magnetic field dependence
in the vortex state of NbSe$_2$. The linear term is considerably
weaker than that determined previously for YBa$_2$Cu$_3$O$_{6.95}$. 
Also, we find that the
vortex-core radius $\rho_0$ which shrinks with increasing field
is not adequately described by the dirty-limit microscopic theory,
but does obey a simple phenomenological equation involving
the intervortex spacing. Our results imply that the conventional GL equations
with field-independent length scales, are not applicable
deep in the superconducting state.

\begin{center} 
\rule{0.3\textwidth}{0.25pt} 
\end{center}  
We are grateful to A.~A.~Golubov and U.~Hartmann
for making available their computer code and
many helpful discussions---as well as
Syd Kreitzman, Curtis Ballard 
and Mel Good for technical assistance.  
This work is supported by NSERC of Canada and
grant NSF-(DMR-95-10453,10454).
       
 
\end{document}